\newif\ifonecol 
\onecolfalse 

\ifonecol
\documentclass[journal,comsoc,12pt,onecolumn]{IEEEtran}
\else
\documentclass[journal,comsoc]{IEEEtran}
\fi

\newlength{\figurewidth}
\ifonecol
\else
\fi

\ifonecol
\usepackage{setspace}
\fi

\usepackage[T1]{fontenc}
\usepackage{color}

\ifCLASSINFOpdf
\else
\fi
\usepackage{amsmath}
\usepackage[cmintegrals]{newtxmath}
\usepackage{mathtools}

\DeclarePairedDelimiter\floor{\lfloor}{\rfloor}

\newcommand{\MYfooter}{\smash{\scriptsize
\hfil\parbox[t][\height][t]{\textwidth}{\centering
\copyright 2020 IEEE. Personal use of this material is permitted. Permission from IEEE must be obtained for all other uses, including reprinting/republishing this material for advertising or promotional purposes, collecting new collected works for resale or redistribution to servers or lists, or reuse of any copyrighted component of this work in other works. DOI: 10.1109/JSYST.2020.2986680}\hfil\hbox{}}}

\makeatletter

\def\ps@IEEEtitlepagestyle{%
\def\@oddfoot{\MYfooter}%
\def\@evenfoot{\MYfooter}}

\makeatother

\begin{document}
%
\title{Performance Analysis of QAM-MPPM in Turbulence-Free FSO Channels: Accurate Derivations and Practical Approximations}
%
%
%

\author{Francisco J. Escribano,~\IEEEmembership{Senior Member,~IEEE,} and Alexandre Wagemakers%
\thanks{Francisco J. Escribano is with the Signal Theory and Communications Department, Universidad de Alcal\'{a}, 28805 Alcal\'{a} de Henares, Spain (email: francisco.escribano@ieee.org).}%
\thanks{Alexandre Wagemakers is with the Nonlinear Dynamics and Chaos Theory Group, Universidad Rey Juan Carlos, 28933 M\'{o}stoles, Spain (email: alexandre.wagemakers@urjc.es).}}%

%
%

\markboth{Journal of \LaTeX\ Class Files,~Vol.~14, No.~8, August~2015}%
{Shell \MakeLowercase{\textit{et al.}}: Bare Demo of IEEEtran.cls for IEEE Communications Society Journals}
%



\maketitle

\begin{abstract}
Following the trends of index modulated (IM) techniques applied to optical communications, several new waveform proposals have been made, aiming at conveying a higher density of information by driving different signal properties. One of these proposals mixes multi-pulse pulse-position modulation (MPPM) and quadrature amplitude modulation (QAM). We present here a new way to demodulate it, and, for the non-turbulent free space optical (FSO) channel, we provide accurate analytical expressions for its error probabilities, both in the case of the traditional and the new detector. We also provide simplified expressions for the estimation of the error probabilities. We show that the new detector offers a gain of some tenths of dB in signal-to-noise ratio with respect to the previously defined one, and that our error probability estimations are more accurate than previously published results. To the best of our knowledge, this work is the first to provide simulation results validating the study of the error probability performance of QAM-MPPM.
\end{abstract}


%
\IEEEpeerreviewmaketitle

\section{Introduction}
%
%
%
%
\IEEEPARstart{I}{n} the latest times, we have been witnessing an increasing interest in new ways to foster the efficiency of digital modulations, due to the prospective demands behind the 5G standarization process, and of the complementary wireless technologies that strive to adapt and survive \cite{8485317}. As an alternative for enhanced designs of the PHY, the concept of index modulation (IM) is gaining momentum \cite{8004416}. Very roughly, the idea behind the IM technique relies in the exploitation of some of the characteristics of the signals or systems involved in a communication setup, in a way where extra information can be carried over, codified in the active communication infrastructure or through specifically chosen parameters.

The initial developments in this field were focused on the transmission frontend, and were specifically related to multiple-input multiple-output (MIMO) setups, where the pattern of active antennas was driven to convey extra information, alongside the usual modulated signals. This was the origin of the so-called spatial modulation (SM) \cite{4382913}, space-time-frequency shift keying (STFS) \cite{5688440}, and other related developments \cite{6166339,6678765}. These ideas were extended to the multiple possible choices of active subcarriers in orthogonal frequency division multiplexing (OFDM), giving rise to OFDM-IM alternatives \cite{6587554}. The IM principles have also been applied to spread spectrum modulation, and code index modulation-spread spectrum (CIM-SS) has been proposed as another index-based modulated system with enhanced capabilities \cite{CIMTVT}.

However, currently envisaged new PHY developments are not exclusively bound to RF: they are also being addressed for optical wireless communications (OWC). The idea is that the usage of light can be a complementary technology apt for the smallest scale deployments, so as to alleviate the scarcity of RF spectrum and face the growing interference limitation concern. The main assets of light are its localized and non-penetrative characteristics. Therefore, the same scenarios exploited in RF about IM have been adapted for OWC, where the stress has been traditionally put on multicarrier applications (OFDM-based setups) and the usage of multiple transmitters and receivers (MIMO-based setups) \cite{7915761,MAO201737,8315127}.

The idea to design IM systems well suited for OWC is also encompassing proposals that go beyond MIMO and OFDM. In single carrier communications there are proposals trying to exploit other additional features of the transmitter/receiver infrastructure. For example, a system has been proposed to jointly use pulse-position modulation (PPM) or frequency shift keying (FSK), while driving the phase or the polarization of the coherent light signal, thus building a compound symbol carrying information along diverse dimensions \cite{Liu:11}. Under the same perspective, optical space modulation (OSM) systems have been proposed, namely, optical space shift keying (OSSK) and spatial pulse position modulation (SPPM) \cite{6241395}. These two schemes constitute appealing solutions for pulse-based OSM systems \cite{7858134}.

Some related proposals rely on using multi-pulse PPM (MPPM) and other properties of the light pulses, like the frequency of their intensity fluctuations \cite{8720057}. If the phase and amplitude of the waveform during the active slots is conveniently driven, it is possible to design a quadrature amplitude modulation (QAM) MPPM system \cite{6876375}. After its initial proposal, such QAM-MPPM system has been studied under different optical channel scenarios and conditions \cite{7858135,7067350,KHALLAF201841}. In the mentioned works, the QAM-MPPM waveform proposed is demodulated using the same metrics for the MPPM and for the QAM part, while the formulas derived for the estimation of the error probabilities were not validated through simulation.

In the present work, we propose another way to demodulate the QAM-MPPM signal, where the detection of the MPPM symbol part relies on an independent set of metrics with respect to the metrics required for the detection of the QAM symbols. The analysis and the simulation results will show that this alternative turns out to be better performing. On the other hand, we will provide exhaustive derivations of the average symbol and bit error probabilities for both kinds of detectors (the traditional one and our alternative), in the case of the non-turbulent free-space optical (FSO) channel. These derivations will lead to practical expressions to estimate the error probabilities that will be validated through simulation, and that will show to be more accurate than the original formulas in previously published works \cite{7858134,6876375}.

To the best of our knowledge, this is the first time that the QAM-MPPM error probability estimations are verified through simulations, and that an exhaustive theoretical derivation of the corresponding expressions is explicitly performed. We are confident that this will help to develop more accurate analytical results to characterize the behavior of QAM-MPPM in more elaborated optical channels. According to all this, we can summarize the main contributions of our paper as follows:
\begin{itemize}
 \item A new detection method for QAM-MPPM.
 \item An analysis of the complexity of both detectors.
 \item A thorough analysis of the average symbol and bit error probabilities for both detectors.
 \item The proposal of useful approximations to actually calculate said probabilities.
 \item The validation of the analytical results by simulation.
\end{itemize}

In Section \ref{model} we review the model for the traditional QAM-MPPM system, and characterize the detectors. In Section \ref{complexity} we compare the complexity of the different detectors. In Section \ref{analysis}, we exhaustively analyze the performance of QAM-MPPM, and derive almost exact expressions for the average symbol and bit error probabilities. In Section \ref{approx} we derive approximations to render usable the expressions calculated in Section \ref{analysis}. In Section \ref{results}, we present simulation results, and validate the tightness of the error probability approximations previously derived. Section \ref{conclusions} is devoted to the conclusions.

%
%

\section{System model}
\label{model}

In our system description, we are going to follow the ideas of the system proposed in \cite{6876375}, but we fully review all the details here to avoid any ambiguity that may prevent the correct understanding of the ensuing developments. Our aim is that all the results can be reproduced by the interested reader without any trouble. As the source of information, we consider an equiprobable binary source of information that produce an independent and identically distributed (i.i.d.) bit sequence that feeds the QAM-MPPM modulator. This modulator divides the signal frame period $T$ into $N$ equal slots of duration $T_s=T/N$. In each frame period, only $1\leq w \leq N$ slots would be active, following an MPPM pattern \cite {Hamkins2005MultipulsePM}, \cite{Caplan2007}. The MPPM symbol is defined by an $N$-dimensional vector $\mathbf{B}$, belonging to the set
\begin{equation}
 \mathcal{S}_{\mathrm{MPPM}}=\left\{ \mathbf{B} \in \{0,1\}^N : \sum_{k=0}^{N-1} B_k=w\right\}.
\end{equation}
The component $B_k$ is $0$ if the slot is not active (non-signal slot), and $1$ if it is active (signal slot). The number of bits carried over per MPPM symbol will be $q_{\mathrm{MPPM}}=\floor*{\log_2 {\binom{N}{w}}}$, which is maximum for $w=\floor*{N/2}$. To send the corresponding information codified in the MPPM symbol, we only use up to $2^{q_{\mathrm{MPPM}}} \leq \binom{N}{w}$ MPPM symbols from the set $\mathcal{S}_{\mathrm{MPPM}}$: we may denote the expurgated MPPM symbol set containing the selected patterns as $\mathcal{S}_{\mathrm{MPPM}}^* \subset \mathcal{S}_{\mathrm{MPPM}}$.

As described in \cite{6876375}, during each signal slot a QAM symbol is inserted, so that the waveform in the electrical domain is
\ifonecol
\begin{equation}
 \label{waveform}
 s\left( t \right)=A\sum\limits_{k=0}^{N-1} B_k p\left(\frac{t-kT_s}{T_s} \right) \biggl[ 1 + m \left( A_k^I \cos \left( 2 \pi f_c t \right) + A_k^Q  \sin\left( 2 \pi f_c t \right) \right)\biggr],
\end{equation}
\else
\begin{eqnarray}
\label{waveform}
& s\left( t \right)=A\sum\limits_{k=0}^{N-1} B_k p\left(\frac{t-kT_s}{T_s} \right) \biggl[ 1 + m  \biggr. \nonumber \\
&  \biggl.  \cdot \left( A_k^I \cos \left( 2 \pi f_c t \right) + A_k^Q  \sin\left( 2 \pi f_c t \right) \right)\biggr],
\end{eqnarray}
\fi
where $p\left(t\right)$ is the unit-duration unit-amplitude rectangular pulse, $B_k$ is the $k-$th component of vector $\mathbf{B}$, $A$ is an amplitude factor, $0 < m \leq 1$ is a modulation index, $f_c=n_c/T_s$ is the carrier frequency ($n_c > 1$, integer), and
\begin{equation}
\left( A_k^I, A_k^Q\right)=\left\{ \begin{array}{ll} \left(0,0\right), & B_k=0 \\ \left(s_{i_k}^I,s_{i_k}^Q\right), & B_k=1 \end{array} \right. ,
\end{equation}
where $\mathbf{s}_i=\left(s_i^I,s_i^Q\right) \in \mathcal{S}_{\mathrm{QAM}}$ is a QAM symbol, $i=0,\cdots,M_Q-1$, and $\mathcal{S}_{\mathrm{QAM}}$ is the QAM symbol set, with $M_Q$ elements. Defining $n_Q=\log_2\left(M_Q\right)$, the number of information bits per QAM-MPPM symbol is
\begin{equation}
 q_{\mathrm{QAM-MPPM}}=q_{\mathrm{MPPM}} + q_{\mathrm{QAM}}=\floor*{\log_2 {\binom{N}{w}}} + w \, n_Q.
\end{equation}
We have considered $M_Q\geq 4$, square-QAM constellations for even $n_Q$, and cross-QAM constellations for odd $n_Q$ (with the exception of $n_Q=3$, where it is rectangular). We also consider gray coding and QAM constellations normalized in energy, so that $E_{\mathrm{QAM}}=\mathrm{E}\left[\| \mathbf{s}_i\|_2^2\right]=1$.

The electrical waveform of equation \eqref{waveform} will linearly drive the light intensity fluctuations of a light source (LED or laser). To avoid clipping, the value of the modulation index $m$ should be set so that $s\left(t\right) \geq 0$. After travelling through a turbulence-free FSO channel, the light intensity fluctuations produced by the light source will hit a photodiode (PD), which will produce a received electrical current waveform
\ifonecol
\begin{equation}
 \label{rwaveform}
 r\left( t \right)=I_{ph}\sum\limits_{k=0}^{N-1} B_k p\left(\frac{t-kT_s}{T_s} \right) \biggl[ 1 + m \left( A_k^I \cos \left( 2 \pi f_c t \right) + A_k^Q  \sin\left( 2 \pi f_c t \right) \right)\biggr] +z\left( t \right),
\end{equation}
\else
\begin{eqnarray}
 \label{rwaveform}
& r\left( t \right)=I_{ph}\sum\limits_{k=0}^{N-1} B_k p\left(\frac{t-kT_s}{T_s} \right) \biggl[ 1 + m  \biggr. \nonumber \\
&  \biggl.  \cdot \left( A_k^I \cos \left( 2 \pi f_c t \right) + A_k^Q  \sin\left( 2 \pi f_c t \right) \right)\biggr] +z\left( t \right),
\end{eqnarray}
\fi
where $I_{ph}$ is the instantaneous PD photocurrent, and $z\left( t \right)$ is an instance of additive white Gaussian noise with power spectral density $N_0/2$. Without loss of generality, ignoring the channel attenuation factor, and the proportional conversion factor between the electrical amplitude and the intensity fluctuations of the light source, the current $I_{ph}$ can be written as
\begin{equation}
 I_{ph}=A \mathcal{R} G,
\end{equation}
where $G$ is the optical channel gain and $\mathcal{R}$ is the responsivity of the PD. The optical channel gain is constant in the case of the turbulence-free FSO channels, and $I_{ph}$ will be therefore considered as a constant value from now on. Notice that we do not consider any dispersive phenomena in the optical channel.

The average received optical power can be calculated as
\begin{equation}
 \label{arop}
 P_{opt}=\frac{I_{DC}}{\mathcal{R}}=\frac{w}{N} \frac{I_{ph}}{\mathcal{R}},
\end{equation}
where
\begin{equation}
 I_{DC}=\frac{w}{N} I_{ph}
\end{equation}
is the DC value of the PD photocurrent for the signal part. The average received symbol energy can be calculated as
\begin{equation}
 \label{Es}
 E_{s,\mathrm{QAM-MPPM}}=w T_s I_{ph}^2 \left(1+\frac{m^2}{2} \right),
\end{equation}
where we have taken into account that the QAM constellation is normalized in energy.  In the signal slots, the average received QAM symbol energy can be written as
\begin{equation}
 \label{EsN0_QAM}
 E_{s,\mathrm{QAM}}=T_s I_{ph}^2 \frac{m^2}{2}.
\end{equation}

For the noise $z\left(t\right)$, we choose a standard model \cite{RoFT02,AOL04}, where the unilateral power spectral density of the noise for the optical receiver can be calculated as
\begin{equation}
\label{N0}
 N_0=\frac{4 k_B T F}{R_L} + 2 \left| q \right| I_{DC} + (RIN) I_{DC}^2 ,
\end{equation}
where $k_B$ is the Bolztmann constant, $T$ is the reference absolute temperature, $F$ is the receiver electronics noise factor, $R_L$ is the PD load resistor, $q$ is the electron charge, and $(RIN)$ is the relative-intensity noise factor. The first term on the RHS is the thermal noise, the second the shot noise, and the third, the relative-intensity noise (RIN).

\subsection{Common metrics detector}

As a first alternative for demodulation, we consider the proposal of \cite{6876375}. For each time slot, $k=0,\cdots,N-1$, resorting to the known principles of the correlator detector for QAM in the signal space framework, the demodulation process will calculate I/Q detected values from \eqref{rwaveform} as
\begin{eqnarray}
 \label{detected}
 & r_k^I=\displaystyle\int_{k T_s}^{\left(k+1\right)T_s} r\left( t \right) \sqrt{\frac{2}{T_s}} \cos\left(2 \pi f_c t\right) dt, \nonumber \\
 & r_k^Q=\displaystyle\int_{k T_s}^{\left(k+1\right)T_s} r\left( t \right) \sqrt{\frac{2}{T_s}} \sin\left(2 \pi f_c t\right) dt,
\end{eqnarray}
where coherent detection is required in the electrical domain. As a result, we have
\begin{eqnarray}
\label{stat_QAM_MPPM}
 r_k^I=\sqrt{\frac{T_s}{2}} I_{ph} B_k m A_k^I + n_k^I; \, \, 
 r_k^Q=\sqrt{\frac{T_s}{2}} I_{ph} B_k m A_k^Q + n_k^Q,
\end{eqnarray}
where $n_k^I$ and $n_k^Q$ are independent zero-mean Gaussian random variables (RVs) with variance $\sigma_n^2=N_0/2$. As done in \cite{6876375}, the MPPM symbol part is detected using the metric
\begin{equation}
 \label{metric}
 X_k = | r_k^I |^2 + | r_k^Q |^2,
\end{equation}
which is a measurement of the detected power of the QAM symbol received in each signal slot. According to the maximum likelihood (ML) rule of \cite{Hamkins2005MultipulsePM}, the values of $0\leq X_k$ can be sorted from highest to lowest, and the first $w$ values will serve to identify the $w$ hypothetical signal slots. As normally $\log_2\binom{N}{w}$ is not an integer, we have to consider two cases. If the resulting MPPM pattern belongs to said set, the corresponding $q_{\mathrm{MPPM}}$ bits can be directly demodulated, according to the chosen mapping. Otherwise, we select the closest MPPM symbol in $\mathcal{S}_{\mathrm{MPPM}}^*$ as appropriate candidate; i.e. if the detected symbol is $\mathbf{B} \notin \mathcal{S}_{\mathrm{MPPM}}^*$, we choose $\mathbf{B}' \in  \mathcal{S}_{\mathrm{MPPM}}^*$, so that
\begin{equation}
 \mathbf{B}' = \arg \min_{\scriptscriptstyle \mathbf{B}^* \in  \mathcal{S}_{\mathrm{MPPM}}^*} \left\{ \| \mathbf{B} - \mathbf{B}^* \|_2^2 \right\}.
\end{equation}
In the case we have a draw among a number of MPPM symbols, the candidate is chosen randomly among them, in order not to introduce any bias. Finally, the information bits mapped in the QAM symbols are demodulated by applying the standard ML demodulator to the I/Q metrics \eqref{stat_QAM_MPPM} of the $w$ hypothetical signal slots identified in the previous step. Notice that the detection of the MPPM symbol and of the QAM symbols involves using the same statistics \eqref{stat_QAM_MPPM}, hence the denomination of common metrics detector (CMD).

It can be shown that $X_k$ follows a scaled chi-square distribution with two degrees of freedom, which is noncentral for the signal slots, and central for the non-signal slots. If the QAM symbol is $\mathbf{s}_i$ in a given signal slot, we can define
\begin{equation}
 \label{Omega}
 \Omega\left(\mathbf{s}_i\right) = T_s I_{ph}^2 \frac{m^2}{2} \| \mathbf{s}_i \|^2,
\end{equation}
and the corresponding conditional probability density function (pdf) of $X_k$ can be defined as
\begin{equation}
\label{Noncentral}
 f_{sl}\left(x;2,\Omega\left(\mathbf{s}_i\right)\right)=\frac{1}{2 \sigma_n^2} \mathrm{e}^{ - \frac{x+\Omega\left(\mathbf{s}_i\right)}{2\sigma_n^2}} \mathrm{I}_0 \left(\frac{\sqrt{x \Omega\left(\mathbf{s}_i\right)}}{\sigma_n^2}\right),
\end{equation}
where $\mathrm{I}_{v}\left(x\right)$ is the $v$-th order modified Bessel function of the first kind. The pdf of $X_k$ for the non-signal slots is
\begin{equation}
\label{Central}
 f_{nsl}\left(x;2\right)=\frac{1}{2 \sigma_n^2} \mathrm{e}^{-\frac{x}{2 \sigma_n^2}}.
\end{equation}
The cumulative distribution functions (cdf's) are, respectively,
\begin{equation}
 \label{distNoncentral}
F_{sl}\left(x;2,\Omega\left(\mathbf{s}_i\right)\right)=1-Q_1\left(\frac{\sqrt{\Omega\left(\mathbf{s}_i\right)}}{\sigma_n},\frac{\sqrt{x}}{\sigma_n}\right) ,
\end{equation}
where $Q_1\left(\cdot,\cdot\right)$ is the first order Marcum-Q function \cite{Pro95}, and
\begin{equation}
 \label{distCentral}
 F_{nsl}\left(x;2\right)=1-\mathrm{e}^{-\frac{x}{2 \sigma_n^2}}.
\end{equation}

\subsection{Independent metrics detector}

\begin{figure*}
\begin{center}
  \includegraphics[width=0.84\textwidth]{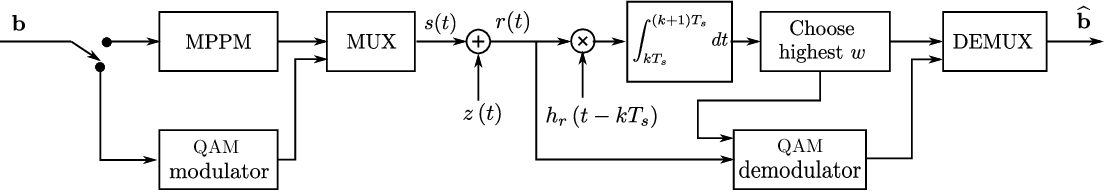}
  \caption{Schematic block of the QAM-MPPM modulator, and the corresponding demodulator for the IMD case.}\label{Fig1}
\end{center}
\vspace*{-0.5cm}
\end{figure*}
In this second alternative, the MPPM symbol will be detected by resorting to a metric independent from \eqref{stat_QAM_MPPM}. In a first stage, we apply the receiver based on the matched filter detector for the rectangular pulse shape, namely
\begin{equation}\label{r_k_MPPM_int}
 r_k=\int_{kT_s}^{\left(k+1\right)T_s} r\left(t\right)  h_r\left(t-kT_s\right) dt,
\end{equation}
where $h_r\left(t\right)=\frac{1}{\sqrt{T_s}}p\left(\frac{t}{T_s}\right)$ is the normalized rectangular pulse receiver filter. Under these conditions, it is easy to verify that
\begin{equation}
\label{r_k_MPPM}
 r_k= \sqrt{T_s} I_{ph} B_k + n_k,
\end{equation}
where $n_k$ is a zero-mean Gaussian RV with variance $\sigma_n^2=N_0/2$. Now we define the metric $X_k=r_k$ (whose support is $-\infty < X_k < \infty$), and detect the signal slots by sorting these values from highest to lowest, according to the already mentioned ML rule of \cite{Hamkins2005MultipulsePM}: the $w$ highest values will determine the hypothetical signal slots. If the detected MPPM pattern does not belong to $\mathcal{S}_{\mathrm{MPPM}}^*$, we apply the same criterion as detailed in the previous type of demodulator. Just as before, once the hypothetical $w$ signal slots have been identified, the standard QAM ML detection process is applied to the I/Q metrics of \eqref{stat_QAM_MPPM}. Notice that we have two different correlation stages here: one to obtain metrics \eqref{r_k_MPPM}, and one to obtain metrics \eqref{stat_QAM_MPPM}. However, this does not suppose a much more complex detector than the previous one. The whole system is represented schematically in Fig. \ref{Fig1} for the IMD.

In this situation, the pdf's of the new RV $X_k$ for the signal and non-signal slots are
\begin{equation}
 \label{pdf_sl_IMD}
 f_{sl}\left(x\right) = \frac{1}{\sqrt{2\pi}\sigma_n} \mathrm{e}^{-\frac{\left(x - \sqrt{T_s} I_{ph}\right)^2}{2 \sigma_n^2}},
\end{equation}
and
\begin{equation}
 \label{pdf_nsl_IMD}
 f_{nsl}\left(x\right) = \frac{1}{\sqrt{2\pi}\sigma_n} \mathrm{e}^{-\frac{x^2}{2 \sigma_n^2}},
\end{equation}
respectively. Their corresponding conditional cdf's are
\begin{equation}
 \label{cdf_sl_IMD}
 F_{sl}\left(x\right) = 1- \frac{1}{2} \mathrm{erfc} \left( \frac{x - \sqrt{T_s} I_{ph}}{\sqrt{2 \sigma_n^2}} \right),
\end{equation}
and
\begin{equation}
 \label{cdf_nsl_IMD}
 F_{nsl}\left(x\right) = 1- \frac{1}{2} \mathrm{erfc} \left( \frac{x}{\sqrt{2 \sigma_n^2}} \right),
\end{equation}
respectively; $\mathrm{erfc}\left(\cdot\right)$ is the complementary error function.

Notice that MPPM and QAM are demodulated based on unrelated statistics, given that the QAM symbol part is cancelled out in \eqref{r_k_MPPM_int}: the MPPM part is detected using the DC value of each slot, while the QAM part is detected using the I/Q coherent demodulator. In particular, metrics \eqref{stat_QAM_MPPM} and \eqref{r_k_MPPM} are independent under the hypothesis that a particular QAM-MPPM symbol has been sent, hence the denomination independent metrics detector (IMD). It is worth stressing the fact that the channel models determined by the receiver statistics in any of the cases (CMD and IMD) constitute instances of discrete-input continuous-output memoryless channels (DCMC).

\section{Complexity Analysis}
\label{complexity}

We now examine the system from the point of view of the computational complexity of the detector. Since the detection process will depend on the chosen metrics, we will detail the analysis separately for each detector. We will assume for both detectors that the signal $r(t)$ is sampled in order to be processed in the digital domain. For a baseband signal with bandwidth equal to $1/T_s$, the minimal number of samples per slot is $N_s=2$ if we want to avoid aliasing, according to the Nyquist-Shannon sampling theorem. The following analysis is not based on a specific implementation, it is rather intended to compare the complexity of the two detectors in fair conditions.

The CMD starts with a computation of the QAM metrics \eqref{stat_QAM_MPPM} at every slot. The filters for the QAM detection have a complexity of $\mathcal{O}\left(N_s\right)$ for each I/Q component. Afterwards, the sorting algorithm for the decision of the most probable QAM symbol has a complexity of $\mathcal{O}\left(M_Q\right)$. The detection of the MPPM symbol can be decomposed in two parts. The first step is to sort the energy of the signal slots using the information of \eqref{metric}, and only the first $w$ slots are retained. This search task can be done with a heap structure with  complexity $\mathcal{O}\left(N\log w\right)$. The second step is to match the MPPM codeword to its corresponding bit sequence. With a LUT that include all possible $\binom{N}{w}$ MPPM patterns, this can be done in constant time. However, for specific values of $N$ and $w$ the LUT size may explode and more efficient techniques are needed \cite{siyu2009,liu2012}. According to these ideas, the computational cost of the CMD based detector is summarized in Table \ref{tab_complexity}.

\begin{table*}
\begin{center}
\begin{tabular}{|l|ll|ll|}
\hline
Detector & \multicolumn{2}{|c|}{CMD} & \multicolumn{2}{|c|}{IMD}  \\
\hline
  & & & Input matched filter:   & $\mathcal{O}\left(N N_s\right)$\\
  & QAM metrics: & $\mathcal{O}\left(2 N N_s \right)$ & QAM metrics: & $\mathcal{O}\left(2 w N_s  \right)$\\
  & QAM sorting: & $\mathcal{O}\left(N M_Q\right)$& QAM sorting: & $\mathcal{O}\left(w M_Q\right)$ \\
  & MPPM : & $\mathcal{O}\left(N\log w\right) + \mathcal{O}\left(1\right)$ &MPPM : & $\mathcal{O}\left(N\log w\right) + \mathcal{O}\left(1\right)$ \\
\hline
Total & \multicolumn{2}{|c|}{$2 N N_s  + N M_Q + N\log w$} & \multicolumn{2}{|c|}{ $N_s (N + 2 w )  + w M_Q + N\log w$}    \\
\hline
\end{tabular}
\end{center}
\vspace*{-0.1cm}
\caption{\label{tab_complexity}Computational cost for the calculation of the IMD and CMD metrics.}
\vspace*{-0.4cm}
\end{table*}
The IMD based detection starts with the computation of the metrics at the output of the matched filter \eqref{r_k_MPPM}, whose complexity is $\mathcal{O}\left(N N_s\right)$. These metrics are sorted with complexity $\mathcal{O}\left(N\log w\right)$ in order to select the $w$ signal slots of the MPPM symbol part. At this point, the detection of the MPPM symbol is identical to the CMD case. Independently, the QAM metrics \eqref{stat_QAM_MPPM} are computed and sorted exclussively over the $w$ previously selected slots. The complexity per symbol and per I/Q component is $\mathcal{O}\left(N_s\right) + \mathcal{O}\left(M_Q\right)$. The total cost for this case is reported in Table \ref{tab_complexity}.

It is interesting to compare the total number of operations for each implementation. For this purpose, we consider a naive implementation of the detector with the possible minimum number of operations. In Table \ref{tab_complexity}, the total number of operations per symbol can be seen for each metrics set. We can approximate the possible gain in complexity from the application of the IMD face to the CMD using the ratio
\begin{equation}
G = \frac{N_s (N+ 2 w) +w M_Q + N \log w}{ 2 N N_s+ N M_Q  + N \log w}.
\end{equation}
In practical cases, this gain is roughly $w/N$, the ratio of signal slots against the total number of slots per frame. This makes sense since the QAM metrics in the case of the IMD are only computed for the detected signal slots, and most of the detection time is spent on the QAM symbol part.

\section{Error probability analysis}
\label{analysis}


\subsection{Average symbol error probability for the CMD}

A QAM-MPPM symbol is defined by the specific set of $w$ QAM symbols from the set $\mathcal{S}_{\mathrm{QAM}}$, and by the specific MPPM pattern $\mathbf{B} \in \mathcal{S}_{\mathrm{MPPM}}^*$. We define as $\mathcal{P}_w^{M_Q}$ the set of all the permutations with repetition $\mathbf{I}=\left\{i_0, \cdots, i_{w-1}\right\}$  of $w$ different indexes $i_j$ taking values in $0,\cdots,M_q-1$. Given an element $\mathbf{I} \in \mathcal{P}_w^{M_Q}$, we may denote the corresponding $w$ QAM symbols in a specific QAM-MPPM symbol as $\left\{\mathbf{s}_{i_j}\right\}_{\mathbf{I}} \in \left(\mathcal{S}_{\mathrm{QAM}}\right)^w$.
It is clear that the number of elements in $\mathcal{P}_w^{M_Q}$ is $\left(M_Q\right)^w=2^{w\cdot n_Q}=2^{q_{\mathrm{QAM}}}$.

\ifonecol
\else
\newcounter{MYtempeqncnt}
\begin{figure*}[!t]
\setcounter{MYtempeqncnt}{\value{equation}}
\setcounter{equation}{34}
\begin{equation}
\label{PcMPPM-pat}
 P_{c,\mathrm{MPPM}}\left(  \left\{\mathbf{s}_{i_l} \right\}_{\mathbf{I}} \right)=\displaystyle\sum_{j=0}^{w-1} \displaystyle\int_{0}^{\infty} f_{sl}\left(x;2,\Omega\left(\mathbf{s}_{i_j}\right)\right) \prod_{l=0, l\neq j}^{w-1} \left( 1 - F_{sl}\left(x;2,\Omega\left(\mathbf{s}_{i_l}\right)\right) \right) F_{nsl}\left(x;2\right)^{N-w} dx.
\end{equation}
\setcounter{equation}{\value{MYtempeqncnt}}
\hrulefill
\end{figure*}
\fi
From these definitions, and taking into account that the input information binary sequence is i.i.d., we may calculate the average symbol error probability of QAM-MPPM as
\ifonecol
\begin{equation}
 \label{averageCMD}
 P_e = \mathrm{E}\left[P_e\left(\left\{ \mathbf{s}_{i_j}\right\}_{\mathbf{I}},\mathbf{B}\right)\right] =\frac{1}{2^{q_{\mathrm{QAM}}}}\frac{1}{2^{q_{\mathrm{MPPM}}}} \sum_{\mathbf{I} \in \mathcal{P}_w^{M_Q}} \sum_{\mathbf{B} \in \mathcal{S}_{\mathrm{MPPM}}^*} P_e\left(\left\{ \mathbf{s}_{i_j}\right\}_{\mathbf{I}},\mathbf{B}\right),
\end{equation}
\else
\begin{eqnarray}
 \label{averageCMD}
 &\displaystyle P_e = \mathrm{E}\left[P_e\left(\left\{ \mathbf{s}_{i_j}\right\}_{\mathbf{I}},\mathbf{B}\right)\right] \\
 &\displaystyle =\frac{1}{2^{q_{\mathrm{QAM}}}}\frac{1}{2^{q_{\mathrm{MPPM}}}} \sum_{\mathbf{I} \in \mathcal{P}_w^{M_Q}} \sum_{\mathbf{B} \in \mathcal{S}_{\mathrm{MPPM}}^*} P_e\left(\left\{ \mathbf{s}_{i_j}\right\}_{\mathbf{I}},\mathbf{B}\right), \nonumber
\end{eqnarray}
\fi
where $P_e\left(\left\{ \mathbf{s}_{i_j}\right\}_{\mathbf{I}},\mathbf{B} \right)$ is the conditional symbol error probability under the hypothesis of having sent a specific QAM-MPPM symbol. This probability could be calculated as one minus the probability of correct detection, which can be factorized as
\begin{equation}
 \label{probcorr}
 P_c\left(\left\{ \mathbf{s}_{i_j}\right\}_{\mathbf{I}},\mathbf{B} \right) = P_{c,\mathrm{MPPM}}\left(\left\{ \mathbf{s}_{i_j}\right\}_{\mathbf{I}},\mathbf{B}\right) \cdot P_{c,\mathrm{QAM}}\left(\left\{ \mathbf{s}_{i_j}\right\}_{\mathbf{I}},\mathbf{B}\right),
\end{equation}
where $P_{c,\mathrm{MPPM}}\left(\left\{ \mathbf{s}_{i_j}\right\}_{\mathbf{I}},\mathbf{B}\right)$ is the probability of correctly detecting the MPPM symbol based on the metrics $X_k$ of equation \eqref{metric}, and $P_{c,\mathrm{QAM}}\left(\left\{ \mathbf{s}_{i_j}\right\}_{\mathbf{I}},\mathbf{B} \right)$ is the conditional probability of correctly demodulating the $w$ QAM symbols using the ML criterion over the I/Q detected values of equations \eqref{stat_QAM_MPPM}, when the MPPM symbol has been correctly detected.

The QAM part, under the hypothesis that the signal slots have been correctly identified, will be independent from the MPPM pattern $\mathbf{B}$, and can be more properly denoted as $P_{c,\mathrm{QAM}}\left(\left\{ \mathbf{s}_{i_j}\right\}_{\mathbf{I}}\right)$. This probability can be calculated as
\begin{equation}
 \label{QAMpart}
 P_{c,\mathrm{QAM}}\left(\left\{ \mathbf{s}_{i_j}\right\}_{\mathbf{I}}\right)=\prod_{j=0}^{w-1} \left( 1 - P_{e,\mathrm{QAM}}\left(\mathbf{s}_{i_j}\right) \right),
\end{equation}
where $P_{e,\mathrm{QAM}}\left(\mathbf{s}_{i_j}\right)$ is the symbol error probability of QAM symbol $\mathbf{s}_{i_j}$ under the hypothesis of ML detection. An approximation for its value will be detailed in Section \ref{approx}.

On the other hand, the derivation of $P_{c,\mathrm{MPPM}}\left(\left\{ \mathbf{s}_{i_j}\right\}_{\mathbf{I}},\mathbf{B}\right)$ is more involved. Resorting to the ideas of \cite{Hamkins2005MultipulsePM} for the case of MPPM in the DCMC, we can calculate it as
\begin{equation}
 \label{MPPMpart}
 P_{c,\mathrm{MPPM}}\left(\left\{ \mathbf{s}_{i_j}\right\}_{\mathbf{I}},\mathbf{B}\right) = \int_{0}^{\infty} p_{sl}\left(x\right) P_{nsl}\left(x\right) dx,
\end{equation}
where $x$ represents the minimum value attained by $X_k$ for the signal slots, $p_{sl}\left(x\right)$ is its pdf, and $P_{nsl}\left(x\right)$ is the cdf of the $N-w$ non-signal slots, representing the probability that their $X_k$ values are lower or equal than $x$. As the RVs $X_k$ are independent from each other, given the hypothesis detailed in the previous section, it is straightforward to see that
\begin{equation}
 P_{nsl}\left(x\right) = F_{nsl}\left(x;2\right)^{N-w},
\end{equation}
where $F_{nsl}\left(x;2\right)$ is given by \eqref{distCentral}. The pdf $p_{sl}\left(x\right)$ can be calculated from its cdf $P_{sl}\left(x\right)$, which, according to \cite{Hamkins2005MultipulsePM}, is
\begin{equation}
 P_{sl}\left(x\right)=1-\prod_{j=0}^{w-1} \left(1-F_{sl}\left(x;2,\Omega\left(\mathbf{s}_{i_j}\right)\right)\right),
\end{equation}
where $F_{sl}\left(x;2,\Omega\left(\mathbf{s}_{i_j}\right)\right)$ is given by \eqref{distNoncentral}. 
Therefore,
\begin{equation}
 \displaystyle p_{sl}\left(x\right) = \sum_{j=0}^{w-1} f_{sl}\left(x;2,\Omega\left(\mathbf{s}_{i_j}\right)\right) \! \prod_{l=0,l\neq j}^{w-1} \! \left(1-F_{sl}\left(x;2,\Omega\left(\mathbf{s}_{i_l}\right)\right)\right), 
\end{equation}
where $f_{sl}\left(x;2,\Omega\left(\mathbf{s}_{i_j}\right)\right)$ is given by \eqref{Noncentral}. As it may be seen, the resulting probability does not depend on the specific MPPM pattern $\mathbf{B}$, and accordingly could be denoted as $P_{c,\mathrm{MPPM}}\left(\left\{ \mathbf{s}_{i_j}\right\}_{\mathbf{I}}\right)$.
\ifonecol
The final expression for the probability of MPPM correct detection can be written as
\begin{equation}
\label{PcMPPM-pat}
 P_{c,\mathrm{MPPM}}\left(  \left\{\mathbf{s}_{i_l} \right\}_{\mathbf{I}} \right)=\displaystyle\sum_{j=0}^{w-1} \displaystyle\int_{0}^{\infty} f_{sl}\left(x;2,\Omega\left(\mathbf{s}_{i_j}\right)\right) \prod_{l=0, l\neq j}^{w-1} \left( 1 - F_{sl}\left(x;2,\Omega\left(\mathbf{s}_{i_l}\right)\right) \right) F_{nsl}\left(x;2\right)^{N-w} dx.
\end{equation}
\else
The final expression for the probability of MPPM correct detection can be seen in equation \eqref{PcMPPM-pat}.
\fi
Notice that in \cite{6876375}, the authors give an expression to calculate the average symbol error probability of MPPM (equation (7) therein), and use it for the calculation of the ensuing QAM-MPPM average error probabilities. It can be seen that (7) in \cite{6876375} and the average over $1-P_{c,\mathrm{MPPM}}\left(  \left\{\mathbf{s}_{i_l} \right\}_{\mathbf{I}} \right)$ (according to equation \eqref{PcMPPM-pat}) lead to dissimilar expressions. As will be shown in the figures devoted to the results for the CMD, our developments fit perfectly the simulated BER, while the results based on (7) in \cite{6876375} provide just a loose upper bound.

Taking all this into account, we can simplify equation \eqref{averageCMD} by cancelling out the dependence on $\mathbf{B}$. On the other hand, we can see from the derived expressions that the particular ordering of the QAM symbols within the QAM-MPPM symbol is irrelevant. Therefore, the probability values will only depend on the specific set of QAM symbols involved: this is represented by the combinations with repetition of $w$ indexes taking values in $0,\cdots,M_Q-1$. If we denote the set of such combinations as $\mathcal{C}_{w}^{M_Q}$, and taking into account that its cardinality is $\binom{M_Q+w-1}{w}$, we may finally rewrite \eqref{averageCMD} as
\ifonecol
\else
\setcounter{MYtempeqncnt}{\value{equation}}
\setcounter{equation}{35}
\fi
\begin{equation}
 \label{simpl_averageCMD}
 \displaystyle P_e = 1 - \frac{1}{\binom{M_Q+w-1}{w}} \sum_{\mathbf{I} \in \mathcal{C}_w^{M_Q}} P_{c,\mathrm{MPPM}}\left(\left\{ \mathbf{s}_{i_j}\right\}_{\mathbf{I}}\right) P_{c,\mathrm{QAM}}\left(\left\{ \mathbf{s}_{i_j}\right\}_{\mathbf{I}}\right).
\end{equation}
Given the expressions \eqref{PcMPPM-pat} and \eqref{QAMpart}, this probability can only be calculated numerically, and may pose stability issues due to the presence of $\mathrm{I}_0\left(\cdot\right)$ in some terms. In Section \ref{approx}, we will address practical methods to calculate $P_e$ for the CMD.

\subsection{Average symbol error probability for the IMD}

As in the case of the CMD, we can calculate the average symbol error probability $P_e$ as the average of equation \eqref{averageCMD}, and the conditional symbol error probability $P_e\left(\left\{ \mathbf{s}_{i_j}\right\}_{\mathbf{I}},\mathbf{B} \right)$ as one minus the conditional probability of correct detection, as factorized in equation \eqref{probcorr}. Since the QAM symbols are detected using the same metrics as before, and under the hypothesis of having correctly identified the MPPM symbol, the conditional probability $P_{c,\mathrm{QAM}}\left(\left\{ \mathbf{s}_{i_j}\right\}_{\mathbf{I}}\right)$ is again given by equation \eqref{QAMpart}.

The derivation of $P_{c,\mathrm{MPPM}}\left(\left\{ \mathbf{s}_{i_j}\right\}_{\mathbf{I}},\mathbf{B}\right)$ is slightly more involved, and has to take into account the new metrics of \eqref{r_k_MPPM}. As in the previous detection mode, the non-signal slots share equal statistics, but, as a difference, this also happens to the signal slots, as the corresponding values of $X_k$ do not depend on the QAM symbols. In this case, it can be seen that the probability of correctly detecting the MPPM symbol part does not depend on $\left\{ \mathbf{s}_{i_j}\right\}_{\mathbf{I}}$, or on the specific MPPM pattern $\mathbf{B}$, and can be more properly written as $P_{c,\mathrm{MPPM}}$. By applying the same criterion of \cite{Hamkins2005MultipulsePM} as before, we can write
\begin{equation}
 \label{MPPMpart_IMD}
 P_{c,\mathrm{MPPM}} = \int_{-\infty}^{\infty} p_{sl}\left(x\right) P_{nsl}\left(x\right) dx,
\end{equation}
where $p_{sl}\left(x\right)$ and $P_{nsl}\left(x\right)$ share the same meaning as in the previous developments, and $x$ is the minimum value attained by the new metrics $X_k$ in the case of the signal slots. Notice that now the integral limit has to be extended from $-\infty$ to $\infty$. It is straightforward to verify that $P_{nsl}\left(x\right)$ can be given as
\begin{equation}
 P_{nsl}\left(x\right) = F_{nsl}\left(x\right)^{N-w},
\end{equation}
where $F_{nsl}\left(x\right)$ is the cdf given in \eqref{cdf_nsl_IMD}. On the other hand, the cdf $P_{sl}\left(x\right)$ would be
\begin{equation}
 P_{sl}\left(x\right) = 1 - \left( 1 - F_{sl}\left(x\right) \right)^w,
\end{equation}
where $F_{sl}\left(x\right)$ is the cdf given in \eqref{cdf_sl_IMD}. Therefore, the probability of correct detection for the MPPM symbol part is
\begin{eqnarray}
\label{probMPPM_IMD}
 P_{c,\mathrm{MPPM}}=\displaystyle w \!\int_{-\infty}^{\infty} \!\! f_{sl}\left(x\right) \left( 1 - F_{sl}\left(x\right) \right)^{w-1} F_{nsl}\left(x\right)^{N-w} dx.
\end{eqnarray}

Using \eqref{QAMpart}, and as the conditional probabilities of correct detection do not depend on $\mathbf{B}$, and the probability $P_{c,\mathrm{MPPM}}$ does not depend on the QAM symbols, we may finally write
\ifonecol
\begin{eqnarray}
\label{Pe_IMD}
 &\displaystyle P_e = 1 -  \frac{P_{c,\mathrm{MPPM}}}{\binom{M_Q+w-1}{w}} \sum_{\mathbf{I} \in \mathcal{C}_w^{M_Q}} P_{c,\mathrm{QAM}}\left(\left\{ \mathbf{s}_{i_j}\right\}_{\mathbf{I}}\right) = 1 -  P_{c,\mathrm{MPPM}} \cdot \mathrm{E}\left[\prod_{j=0}^{w-1} \left( 1 - P_{e,\mathrm{QAM}}\left(\mathbf{s}_{i_j}\right) \right)\right] & \nonumber\\
 &\displaystyle = 1 -  P_{c,\mathrm{MPPM}} \left(1 - P_{e,\mathrm{QAM}} \right)^w, &
\end{eqnarray}
\else
\begin{eqnarray}
\label{Pe_IMD}
 &\displaystyle P_e = 1 -  \frac{P_{c,\mathrm{MPPM}}}{\binom{M_Q+w-1}{w}} \sum_{\mathbf{I} \in \mathcal{C}_w^{M_Q}} P_{c,\mathrm{QAM}}\left(\left\{ \mathbf{s}_{i_j}\right\}_{\mathbf{I}}\right)& \nonumber \\
 &\displaystyle = 1 -  P_{c,\mathrm{MPPM}} \cdot \mathrm{E}\left[\prod_{j=0}^{w-1} \left( 1 - P_{e,\mathrm{QAM}}\left(\mathbf{s}_{i_j}\right) \right)\right] & \nonumber \\
 &\displaystyle = 1 -  P_{c,\mathrm{MPPM}} \left(1 - P_{e,\mathrm{QAM}} \right)^w,
\end{eqnarray}
\fi
where $P_{e,\mathrm{QAM}}$ is the average symbol error probability of  QAM \cite{Pro95}, calculated using $E_{s,\mathrm{QAM}}$ as defined in \eqref{EsN0_QAM}.

\subsection{Average bit error probability for the CMD}

\ifonecol
\else
\begin{figure*}[!t]
\setcounter{MYtempeqncnt}{\value{equation}}
\setcounter{equation}{50}
\begin{eqnarray}
 \label{Pbtot_CMD}
 &\displaystyle P_b = \frac{1}{q_{\mathrm{QAM-MPPM}}} \left[ \frac{1}{\binom{M_Q+w-1}{w}} \! \sum_{\mathbf{I}\in \mathcal{C}_{w}^{M_Q}} \!\!\! P_{c,\mathrm{MPPM}}\left(\left\{ \mathbf{s}_{i_j}\right\}_{\mathbf{I}}\right) \! \sum_{j=0}^{w-1} \! P_{e,\mathrm{QAM}}\left(\mathbf{s}_{i_j}\right) 
+ \! \frac{2^{q_{\mathrm{MPPM}}-1}}{2^{q_{\mathrm{MPPM}}}-1} \frac{q_{\mathrm{MPPM}}}{\binom{M_Q+w-1}{w}} \! \sum_{\mathbf{I}\in \mathcal{C}_{w}^{M_Q}} \!\!\!\left( 1- P_{c,\mathrm{MPPM}}\left(\left\{ \mathbf{s}_{i_j}\right\}_{\mathbf{I}}\right) \right) \right. \nonumber \\
&\displaystyle \left. + \frac{1}{\binom{M_Q+w-1}{w}} \sum_{\mathbf{I}\in \mathcal{C}_{w}^{M_Q}} \left( 1- P_{c,\mathrm{MPPM}}\left(\left\{ \mathbf{s}_{i_j}\right\}_{\mathbf{I}}\right) \right) \left(\sum_{l=1}^{\min\left(w,N-w\right)} K_l \left( \left(w-l\right) \frac{1}{w}  \sum_{j=0}^{w-1} P_{e,\mathrm{QAM}}\left(\mathbf{s}_{i_j}\right) + \frac{n_Q}{2} l \right)  \right) \right]  
\end{eqnarray}
\setcounter{equation}{\value{MYtempeqncnt}}
\hrulefill
\end{figure*}
\fi
To calculate the average bit error probability, we can average over the conditional bit error probability, so that
\ifonecol
\begin{equation}
 \label{averageCMD_Pb}
  P_b = \mathrm{E}\left[P_b\left(\left\{ \mathbf{s}_{i_j}\right\}_{\mathbf{I}},\mathbf{B}\right)\right] =\frac{1}{2^{q_{\mathrm{QAM}}}}\frac{1}{2^{q_{\mathrm{MPPM}}}} \sum_{\mathbf{I} \in \mathcal{P}_w^{M_Q}} \sum_{\mathbf{B} \in \mathcal{S}_{\mathrm{MPPM}}^*} P_b\left(\left\{ \mathbf{s}_{i_j}\right\}_{\mathbf{I}},\mathbf{B}\right).
\end{equation}
\else
\begin{eqnarray}
 \label{averageCMD_Pb}
 &\displaystyle P_b = \mathrm{E}\left[P_b\left(\left\{ \mathbf{s}_{i_j}\right\}_{\mathbf{I}},\mathbf{B}\right)\right] \\
 &\displaystyle =\frac{1}{2^{q_{\mathrm{QAM}}}}\frac{1}{2^{q_{\mathrm{MPPM}}}} \sum_{\mathbf{I} \in \mathcal{P}_w^{M_Q}} \sum_{\mathbf{B} \in \mathcal{S}_{\mathrm{MPPM}}^*} P_b\left(\left\{ \mathbf{s}_{i_j}\right\}_{\mathbf{I}},\mathbf{B}\right). \nonumber
\end{eqnarray}
\fi
The probability $P_b\left(\left\{ \mathbf{s}_{i_j}\right\}_{\mathbf{I}},\mathbf{B}\right)$ can be factorized under the mutually exclusive hypothesis of correct and erroneous demodulation of MPPM, so that
\ifonecol
\begin{equation}
 P_b\left(\left\{ \mathbf{s}_{i_j}\right\}_{\mathbf{I}},\mathbf{B}\right) = p_b\left(\left\{ \mathbf{s}_{i_j}\right\}_{\mathbf{I}},\mathbf{B} \big\vert c, \mathrm{MPPM} \right) P_{c,\mathrm{MPPM}}\left(\left\{ \mathbf{s}_{i_j}\right\}_{\mathbf{I}}\right) + p_b\left(\left\{ \mathbf{s}_{i_j}\right\}_{\mathbf{I}},\mathbf{B} \big\vert e, \mathrm{MPPM} \right) \left( 1 -  P_{c,\mathrm{MPPM}}\left(\left\{ \mathbf{s}_{i_j}\right\}_{\mathbf{I}}\right)\right),
\end{equation}
\else
\begin{eqnarray}
 & P_b\left(\left\{ \mathbf{s}_{i_j}\right\}_{\mathbf{I}},\mathbf{B}\right)  & \\
 & = p_b\left(\left\{ \mathbf{s}_{i_j}\right\}_{\mathbf{I}},\mathbf{B} \big\vert c, \mathrm{MPPM} \right) P_{c,\mathrm{MPPM}}\left(\left\{ \mathbf{s}_{i_j}\right\}_{\mathbf{I}}\right) & \nonumber \\
 & + p_b\left(\left\{ \mathbf{s}_{i_j}\right\}_{\mathbf{I}},\mathbf{B} \big\vert e, \mathrm{MPPM} \right) \left( 1 -  P_{c,\mathrm{MPPM}}\left(\left\{ \mathbf{s}_{i_j}\right\}_{\mathbf{I}}\right)\right), & \nonumber
\end{eqnarray}
\fi
where the probability $P_{c,\mathrm{MPPM}}\left(\left\{ \mathbf{s}_{i_j}\right\}_{\mathbf{I}}\right)$ is given in equation \eqref{PcMPPM-pat}, and $p_b\left(\cdot,\cdot \vert \cdot \right)$ is the proportion of erroneous bits under the given hypothesis. The value $p_b\left(\left\{ \mathbf{s}_{i_j}\right\}_{\mathbf{I}},\mathbf{B} \big\vert c, \mathrm{MPPM} \right)$ is the proportion of bits in error in the demodulation of the QAM-MPPM symbol when the MPPM demodulation has correctly identified the signal slots, and only the errors in demodulating the QAM symbols have to be taken into account. The specific value of $\mathbf{B}$ is irrelevant, and only $\left\{\mathbf{s}_{i_j}\right\}_{\mathbf{I}}$ matters. We can calculate it as
\begin{equation}
 p_b\left(\left\{ \mathbf{s}_{i_j}\right\}_{\mathbf{I}},\mathbf{B} \big\vert c, \mathrm{MPPM} \right) = \frac{ne_{\mathrm{QAM}}}{q_{\mathrm{QAM-MPPM}}},
\end{equation}
where $ne_{\mathrm{QAM}}$ is the average number of erroneous bits determined by the QAM symbols. As the mapping from bits to symbols is gray, we can approximate the bit error probability associated to symbol $\mathbf{s}_{i_j}$ as $P_{e,\mathrm{QAM}}\left(\mathbf{s}_{i_j}\right)/n_Q$. This is the proportion of erroneous bits in the QAM symbol, and $n_Q \cdot P_{e,\mathrm{QAM}}\left(\mathbf{s}_{i_j}\right)/n_Q$ will be its contribution to the total. As we have a set of $w$ QAM symbols,
\begin{equation}
 ne_{\mathrm{QAM}} = \sum_{j=0}^{w-1} P_{e,\mathrm{QAM}}\left(\mathbf{s}_{i_j}\right).
\end{equation}

For the complementary hypothesis, we have
\begin{equation}
 p_b\left(\left\{ \mathbf{s}_{i_j}\right\}_{\mathbf{I}},\mathbf{B} \big\vert e, \mathrm{MPPM} \right) = \frac{ne_{\mathrm{MPPM}}+ne_{\mathrm{QAM}}^{sl}+ne_{\mathrm{QAM}}^{nsl}}{q_{\mathrm{QAM-MPPM}}},
\end{equation}
where $ne_{\mathrm{MPPM}}$ is the average number of erroneous bits in the demodulation of MPPM, $ne_{\mathrm{QAM}}^{sl}$ is the average number of erroneous bits in the demodulation of QAM for the proportion of correctly identified signal slots, and $ne_{\mathrm{QAM}}^{nsl}$ is the average number of erroneous bits when applying QAM demodulation to the non-signal slots erroneously identified as signal slots. The estimated proportion of bits affected by an MPPM detection error is $2^{q_{\mathrm{MPPM}}-1}/\left(2^{q_{\mathrm{MPPM}}}-1\right)$ \cite{a029b3b4b02e4b40ac8ca72cc4f7fc8c}, so that the corresponding average number of erroneous bits will be
\begin{equation}
\label{eq_ne_MPPM}
 ne_{\mathrm{MPPM}} = q_{\mathrm{MPPM}} \frac{2^{q_{\mathrm{MPPM}}-1}}{2^{q_{\mathrm{MPPM}}}-1}.
\end{equation}
As previously seen, for $ne^{sl}_{\mathrm{QAM}}$ we have an average number of erroneous bits per QAM symbol of $P_{e,\mathrm{QAM}}\left(\mathbf{s}_{i_j}\right)$, and now we have to take into account the average number of signal slots correctly identified. This can be calculated as \cite{6620996}
\begin{equation}
 \frac{\sum\limits_{l=1}^{\min\left(w,N-w\right)} {\binom{w}{l}} {\binom{N-w}{l}} \left(w-l\right) }{\binom{N}{w}-1} = \sum_{l=1}^{\min\left(w,N-w\right)} K_l \left(w-l\right),
\end{equation}
where the index $l$ is the number of signal slots missed in the detection of MPPM, and we have defined $K_l={\binom{w}{l}} {\binom{N-w}{l}}/\left({\binom{N}{w}-1}\right)$. Consequently,
\begin{equation}
 ne_{\mathrm{QAM}}^{sl} = \sum\limits_{l=1}^{\min\left(w,N-w\right)} K_l \left(w-l\right) \frac{1}{w} \sum_{j=0}^{w-1} P_{e,\mathrm{QAM}}\left(\mathbf{s}_{i_j}\right). 
\end{equation}
In the case of $ne_{\mathrm{QAM}}^{nsl}$, we can make the reasonable assumption that on average half of the bits involved in the demodulation of QAM over a non-signal slot will be in error, so that
\begin{equation}
 ne_{\mathrm{QAM}}^{nsl} =  \frac{n_Q}{2} \sum\limits_{l=1}^{\min\left(w,N-w\right)} K_l \, l.  
\end{equation}
Notice that we are implicitly assuming that all the possible MPPM patterns in $\mathcal{S}_{\mathrm{MPPM}}$, excepting the hypothetical $\mathbf{B}\in \mathcal{S}_{\mathrm{MPPM}}^*$, can be chosen in the demodulation. Therefore, these expressions for $ne_{\mathrm{MPPM}}$,  $ne^{sl}_{\mathrm{QAM}}$  and $ ne^{nsl}_{\mathrm{QAM}}$ will only be exact if $\log_2\binom{N}{w}$ is an integer. As normally this is not the case, the results should be then interpreted as approximations, but, given that the difference between $\log_2\binom{N}{w}$ and $\floor*{\log_2\binom{N}{w}}$ is in practice small, the resulting penalty will not be high.

\ifonecol
As none of the terms involved depend on the specific MPPM symbol $\mathbf{B}$, or on the specific location of the QAM symbols, the average in \eqref{averageCMD_Pb} can be finally written as
\begin{eqnarray}
 \label{Pbtot_CMD}
 &\displaystyle P_b = \frac{1}{q_{\mathrm{QAM-MPPM}}} \left[ \frac{1}{\binom{M_Q+w-1}{w}} \! \sum_{\mathbf{I}\in \mathcal{C}_{w}^{M_Q}} \!\!\! P_{c,\mathrm{MPPM}}\left(\left\{ \mathbf{s}_{i_j}\right\}_{\mathbf{I}}\right) \! \sum_{j=0}^{w-1} \! P_{e,\mathrm{QAM}}\left(\mathbf{s}_{i_j}\right) \right. & \nonumber \\
&\displaystyle \left. + \frac{2^{q_{\mathrm{MPPM}}-1}}{2^{q_{\mathrm{MPPM}}}-1} \frac{q_{\mathrm{MPPM}}}{\binom{M_Q+w-1}{w}} \! \sum_{\mathbf{I}\in \mathcal{C}_{w}^{M_Q}} \!\!\!\left( 1- P_{c,\mathrm{MPPM}}\left(\left\{ \mathbf{s}_{i_j}\right\}_{\mathbf{I}}\right) \right) \right. \nonumber \\
&\displaystyle \left. + \frac{1}{\binom{M_Q+w-1}{w}} \sum_{\mathbf{I}\in \mathcal{C}_{w}^{M_Q}} \left( 1- P_{c,\mathrm{MPPM}}\left(\left\{ \mathbf{s}_{i_j}\right\}_{\mathbf{I}}\right) \right) \left(\sum_{l=1}^{\min\left(w,N-w\right)} K_l \left( \left(w-l\right) \frac{1}{w}  \sum_{j=0}^{w-1} P_{e,\mathrm{QAM}}\left(\mathbf{s}_{i_j}\right) + \frac{n_Q}{2} l \right)  \right) \right]  
\end{eqnarray}
\else
As none of the terms involved depend on the specific MPPM symbol $\mathbf{B}$, or on the specific location of the QAM symbols, the average in \eqref{averageCMD_Pb} can be finally written as shown in \eqref{Pbtot_CMD}.
\fi
This expression is very similar to the one developed in \cite{6620996}, where the modulation used was BPSK. Notice also that the authors of \cite{6876375} provide an expression for the average bit error probability of QAM-MPPM (equation (6) therein) which is different from \eqref{Pbtot_CMD}. Apart from considering separate averages for the MPPM and the QAM error probability parts, they do not take into account the cases where just a fraction of the $w$ slots have been correctly identified. This is a conservative approach that would clearly lead to an upper bound. Therefore, for fairness we have chosen to use \eqref{Pbtot_CMD} to calculate the average bit error probability of all the CMD cases, even when comparing our results with the ones obtained by using (7) in \cite{6876375} for the calculation of the MPPM average symbol error probability part.

\subsection{Average bit error probability for the IMD}

Now equation \eqref{Pbtot_CMD} is still valid, but it admits some simplifications. We know that the probability of correctly detecting MPPM does not depend on the QAM symbols, and $P_{c,\mathrm{MPPM}}\left(\left\{ \mathbf{s}_{i_j}\right\}_{\mathbf{I}}\right)=P_{c,\mathrm{MPPM}}$ as seen in \eqref{probMPPM_IMD}. Therefore, in the corresponding terms of equation \eqref{Pbtot_CMD} we will have just the averaging of the QAM symbol probabilities. For example, in the first term in the RHS of \eqref{Pbtot_CMD}, we arrive at
\ifonecol
\else
\setcounter{MYtempeqncnt}{\value{equation}}
\setcounter{equation}{51}
\fi
\begin{equation}
 \frac{1}{\binom{M_Q+w-1}{w}} \sum_{\mathbf{I}\in \mathcal{C}_{w}^{M_Q}} \sum_{j=0}^{w-1} P_{e,\mathrm{QAM}}\left(\mathbf{s}_{i_j}\right) = w P_{e,\mathrm{QAM}}.
\end{equation}

According to this, the bit error probability simplifies to
\ifonecol
\begin{eqnarray}
 \label{Pbtot_IMD}
 &\displaystyle P_b = \frac{1}{q_{\mathrm{QAM-MPPM}}} \Bigg[ P_{c,\mathrm{MPPM}} w P_{e,\mathrm{QAM}} + \frac{2^{q_{\mathrm{MPPM}}-1}}{2^{q_{\mathrm{MPPM}}}-1} q_{\mathrm{MPPM}} \left( 1- P_{c,\mathrm{MPPM}} \right) & \nonumber \\
 &\displaystyle + \left( 1- P_{c,\mathrm{MPPM}} \right) \! \left( \sum\limits_{l=1}^{\min\left(w,N-w\right)} \!\!\!\!\! K_l \left( \left(w-l\right) P_{e,\mathrm{QAM}} + \frac{n_Q}{2} l  \right)\right) \Bigg].
\end{eqnarray}
\else
\begin{eqnarray}
 \label{Pbtot_IMD}
 &\displaystyle P_b = \frac{1}{q_{\mathrm{QAM-MPPM}}} \Bigg[ P_{c,\mathrm{MPPM}} w P_{e,\mathrm{QAM}} \\
 &\displaystyle + \frac{2^{q_{\mathrm{MPPM}}-1}}{2^{q_{\mathrm{MPPM}}}-1} q_{\mathrm{MPPM}} \left( 1- P_{c,\mathrm{MPPM}} \right) \nonumber \\
 &\displaystyle + \left( 1- P_{c,\mathrm{MPPM}} \right) \! \left( \sum\limits_{l=1}^{\min\left(w,N-w\right)} \!\!\!\!\!\! K_l \left( \left(w-l\right) P_{e,\mathrm{QAM}} + \frac{n_Q}{2} l  \right)\right) \Bigg]. \nonumber
\end{eqnarray}
\fi
Notice that, in general, we cannot simplify equation \eqref{Pbtot_CMD} in the same way for the CMD, since each subset of QAM symbols will lead to a different probability of successfully demodulating the MPPM symbol part $P_{c,\mathrm{MPPM}}\left(\left\{ \mathbf{s}_{i_j}\right\}_{\mathbf{I}}\right)$ (see \eqref{PcMPPM-pat}). Moreover, the individual QAM symbol error probability is in general not the same for all of them. Nevertheless, some approximations may be made to address in practice the calculations of \eqref{simpl_averageCMD}, \eqref{Pe_IMD}, \eqref{Pbtot_CMD} and \eqref{Pbtot_IMD}, as shown in the next Section.

\section{Practical approximations}
\label{approx}

We address here approximations and simplifications that render tractable the task of using the formulas developed in Section \ref{analysis}. In their original proposal and other related works, to the best of our knowledge, the authors of \cite{6876375} do not provide practical ways of using their own formulas.

\subsection{Common metrics detector}

We have two possibilities to perform the calculations.

\subsubsection{Numerical integration approach with joint averages}

The numerical calculation of the integral in \eqref{PcMPPM-pat} faces the problem of the instability of  $\mathrm{I}_0\left(x\right)$. However, in the typical mathematical software packages, it is possible to resort to the scaled $v$-th order modified Bessel function of the first kind
\begin{equation}
 \mathrm{I}_v^S\left(x\right) = \mathrm{I}_v\left(x\right) \mathrm{e}^{-x}.
\end{equation}
Using this, and with a little algebra, the pdf of the signal slots \eqref{Noncentral} can be rewritten as
\begin{equation}
\label{scaled_ver}
 f_{sl}\left(x;2,\Omega\left(\mathbf{s}_i\right)\right)=\frac{1}{2 \sigma_n^2} \mathrm{e}^{ - \frac{\left(\sqrt{x}+\sqrt{\Omega\left(\mathbf{s}_i\right)}\right)^2}{2\sigma_n^2}} \mathrm{I}_0^S\left(\frac{\sqrt{x \Omega\left(\mathbf{s}_i\right)}}{\sigma_n^2}\right),
\end{equation}
for real values of $x$. Using this expression, the resulting integral for equation \eqref{PcMPPM-pat} can be now numerically integrated without stability issues.

In order to perform the required averages over the combinations of $w$ QAM symbols, we need an estimation of the individual symbol error probability. In this case, we can use the union bound (UB) approximation, so that
\begin{equation}
\label{UB_QAM}
  \displaystyle P_{e,\mathrm{QAM}}\left(\mathbf{s}_i\right) \leq \frac{1}{2} \sum_{j=0,j\neq i}^{M_Q-1} \mathrm{erfc}\left( \sqrt{\frac{ T_s I_{ph}^2 m^2 \| \mathbf{s}_i - \mathbf{s}_j \|^2}{16 \sigma_n^2}} \right).
\end{equation}
With these definitions, now we can calculate an approximation to the joint averages in \eqref{simpl_averageCMD} and \eqref{Pbtot_CMD}, and thus to the average symbol and bit error probabilities. Notice that we are not making approximations other than the mentioned ones (ignoring the difference between $\mathcal{S}^*_{\mathrm{MPPM}}$ and $\mathcal{S}_{\mathrm{MPPM}}$, and applying the union bound technique for the symbol error probability of QAM). Due to this, it is expected that the main difference with respect to the true error rates calculated through simulation will not be significant. We will denote this scenario as CMD/JA.

\subsubsection{Numerical integration approach with separate averages}

The previous calculations can take a lot of time, and require intensive memory resources. One possibility to reduce these demands consists in approximating the expectation of the product of functions of $P_{c,\mathrm{MPPM}}\left(\left\{ \mathbf{s}_{i_j}\right\}_{\mathbf{I}}\right)$ and $P_{c,\mathrm{QAM}}\left( \mathbf{s}_i \right)$ (or $P_{e,\mathrm{QAM}}\left( \mathbf{s}_i \right)$) by the product of the corresponding expectations. For example, in the case of the average symbol error probability for QAM-MPPM,
\ifonecol
\begin{equation}
\label{assumption_indep}
  P_e = 1 - \mathrm{E}\left[P_{c,\mathrm{MPPM}}\left(\left\{ \mathbf{s}_{i_j}\right\}_{\mathbf{I}}\right) P_{c,\mathrm{QAM}}\left(\left\{ \mathbf{s}_{i_j}\right\}_{\mathbf{I}}\right) \right] \approx 1 - \mathrm{E}\left[P_{c,\mathrm{MPPM}}\left(\left\{ \mathbf{s}_{i_j}\right\}_{\mathbf{I}}\right)\right] \mathrm{E}\left[P_{c,\mathrm{QAM}}\left(\left\{ \mathbf{s}_{i_j}\right\}_{\mathbf{I}}\right) \right].
\end{equation}
\else
\begin{eqnarray}
\label{assumption_indep}
  & \displaystyle P_e = 1 - \mathrm{E}\left[P_{c,\mathrm{MPPM}}\left(\left\{ \mathbf{s}_{i_j}\right\}_{\mathbf{I}}\right) P_{c,\mathrm{QAM}}\left(\left\{ \mathbf{s}_{i_j}\right\}_{\mathbf{I}}\right) \right]\nonumber \\
  & \displaystyle \approx 1 - \mathrm{E}\left[P_{c,\mathrm{MPPM}}\left(\left\{ \mathbf{s}_{i_j}\right\}_{\mathbf{I}}\right)\right] \mathrm{E}\left[P_{c,\mathrm{QAM}}\left(\left\{ \mathbf{s}_{i_j}\right\}_{\mathbf{I}}\right) \right].
\end{eqnarray}
\fi
This finally leads to the same expression as in the IMD case \eqref{Pe_IMD}, with the appropriate definitions for each expectation. This strategy can be applied to each of the terms of the bit error probability of equation \eqref{Pbtot_CMD} where an average of a product of functions of the conditional probabilities for QAM and MPPM exists. The difference between both approaches (the joint and the separate ones) will be shown to be numerically negligible, but the second alternative will be far less time consuming.

The averages over the expressions containing conditional symbol error probabilities for QAM will lead to the usage of the known expressions for the QAM average symbol error probabilities \cite{Pro95}, as previously seen. The average for $P_{c,\mathrm{MPPM}}\left(\left\{ \mathbf{s}_{i_j}\right\}_{\mathbf{I}}\right)$ can be readily calculated as
\ifonecol
\begin{equation}
\label{Pc_IMD_SA}
 \mathrm{E}\left[P_{c,\mathrm{MPPM}}\left(\left\{ \mathbf{s}_{i_j}\right\}_{\mathbf{I}}\right)\right] = w \int_{0}^{\infty} f_{sl}\left(x;2\right) \left( 1 - F_{sl}\left(x;2\right) \right)^{w-1} F_{nsl}\left(x;2\right)^{N-w} dx,
\end{equation}
\else
\begin{eqnarray}
\label{Pc_IMD_SA}
 & \displaystyle \mathrm{E}\left[P_{c,\mathrm{MPPM}}\left(\left\{ \mathbf{s}_{i_j}\right\}_{\mathbf{I}}\right)\right] \\
 & \displaystyle = w \int_{0}^{\infty} f_{sl}\left(x;2\right) \left( 1 - F_{sl}\left(x;2\right) \right)^{w-1} F_{nsl}\left(x;2\right)^{N-w} dx, \nonumber
\end{eqnarray}
\fi
where $F_{nsl}\left(x;2\right)$ is given by \eqref{distCentral} and
\begin{eqnarray}
 \label{uncondf}
 & \displaystyle f_{sl}\left(x;2\right) = \frac{1}{M_Q} \sum_{i=0}^{M_Q-1} f_{sl}\left(x;2,\Omega\left(\mathbf{s}_i\right)\right), \\
 \label{uncondF}
 & \displaystyle F_{sl}\left(x;2\right) = \frac{1}{M_Q} \sum_{i=0}^{M_Q-1} F_{sl}\left(x;2,\Omega\left(\mathbf{s}_i\right)\right),
\end{eqnarray}
are the unconditional pdf and cdf of the signal slots, respectively. This result takes into account the linearity of the expectation operator, and the independence in the occurrence of the different QAM symbols at each of the signal slots. By using \eqref{scaled_ver}, we may again define integrals that can be calculated numerically without trouble using standard mathematical software. We will denote this scenario as CMD/SA.

\subsection{Independent metrics detector}

We have again two different approaches for the calculations.

\subsubsection{Numerical integration approach}

If we focus on equation \eqref{probMPPM_IMD}, we can see that
\ifonecol
\begin{equation}
 P_{c,\mathrm{MPPM}}= w \sum_{m=0}^{N-w} \binom{N-w}{m}  \frac{\left(-1\right)^m}{\sqrt{2\pi}\sigma_n}  \int_{-\infty}^{\infty} \mathrm{e}^{-\frac{\left(x - \sqrt{T_s} I_{ph}\right)^2}{2 \sigma_n^2}} \left( \frac{1}{2} \mathrm{erfc} \left( \frac{x - \sqrt{T_s} I_{ph}}{\sqrt{2 \sigma_n^2}} \right)\right)^{w-1} \left(\frac{1}{2} \mathrm{erfc} \left( \frac{x}{\sqrt{2 \sigma_n^2}} \right)\right)^{m}  dx,
\end{equation}
\else
\begin{eqnarray}
 & \displaystyle P_{c,\mathrm{MPPM}}= w \sum_{m=0}^{N-w} \binom{N-w}{m}  \frac{\left(-1\right)^m}{\sqrt{2\pi}\sigma_n}  \int_{-\infty}^{\infty} \mathrm{e}^{-\frac{\left(x - \sqrt{T_s} I_{ph}\right)^2}{2 \sigma_n^2}} \nonumber \\
 & \displaystyle \cdot  \left( \frac{1}{2} \mathrm{erfc} \left( \frac{x - \sqrt{T_s} I_{ph}}{\sqrt{2 \sigma_n^2}} \right)\right)^{w-1} \left(\frac{1}{2} \mathrm{erfc} \left( \frac{x}{\sqrt{2 \sigma_n^2}} \right)\right)^{m}  dx,
\end{eqnarray}
\fi
where we have replaced the corresponding probability density and cumulative distribution functions, and developed the binomial corresponding to $F_{nsl}\left(x\right)^{N-w}$. The resulting integrals can be numerically calculated without stability issues. We will denote this scenario as IMD/NI.

\subsubsection{Union bound approach}

On the other hand, as seen in \cite{6241395}, the demodulation method chosen for MPPM is equivalent to finding the MPPM vector $\mathbf{B}$ closest (in the Euclidean distance sense) to $\left(r_0,\cdots,r_{N-1}\right)$, where $r_k$ are the received values of equation \eqref{r_k_MPPM}. In this case, $P_{c,\mathrm{MPPM}}$ can be calculated as one minus the average symbol error probability of MPPM ($P_{e,\mathrm{MPPM}}$), approximated as the UB
\begin{equation}
 \displaystyle P_{e,\mathrm{MPPM}} \leq \frac{1}{2^{q_{\mathrm{MPPM}}+1}} \! \sum_{\scriptscriptstyle\mathbf{B} \in \mathcal{S}_{\mathrm{MPPM}}^{*}} \sideset{}{^{\scriptscriptstyle\mathbf{B}' \neq \mathbf{B}}}\sum_{\hspace{10pt} \scriptscriptstyle\mathbf{B}' \in \mathcal{S}_{\mathrm{MPPM}}^{*}} \!\! \mathrm{erfc}\left(\sqrt{\frac{T_s I_{ph}^2 \| \mathbf{B} - \mathbf{B}' \|_2^2}{8 \sigma_n^2}} \right),
\end{equation}
where we have only taken into account the valid MPPM patterns in the expurgated set $\mathcal{S}_{\mathrm{MPPM}}^*$. We will denote this scenario as IMD/UB.

\section{Simulation results and discussion}
\label{results}

Apart from using simulation results to validate our approaches, we resort to previously published formulas in \cite{6876375} (reused in \cite{7858135}) for QAM-MPPM, in order to assess their relative accuracy. To the best of our knowledge, we are the first to provide this analysis for QAM-MPPM. We will show that our results will be in general tighter. To render usable expression (7) in \cite{6876375} for the CMD, we require density functions corresponding to the signal slots that do not depend on the QAM symbols. The pdf and cdf of the signal slots are denoted there as $p_1\left(\cdot\right)$ and $P_1\left(\cdot\right)$, respectively. As the formulas in \cite{6876375} implicitly assume separate averages (even when originally dealing with the CMD), $p_1\left(\cdot\right)$ should be given by \eqref{uncondf}, and $P_1\left(\cdot\right)$ by \eqref{uncondF}. For the IMD case, $p_1\left(\cdot\right)$ and $P_1\left(\cdot\right)$ should be given by \eqref{pdf_sl_IMD} and \eqref{cdf_sl_IMD}, respectively. The respective scenarios are labeled as CMD/\cite{6876375}, and IMD/\cite{6876375}.
\begin{figure}[htb!]
\centering
\includegraphics[width=\figurewidth]{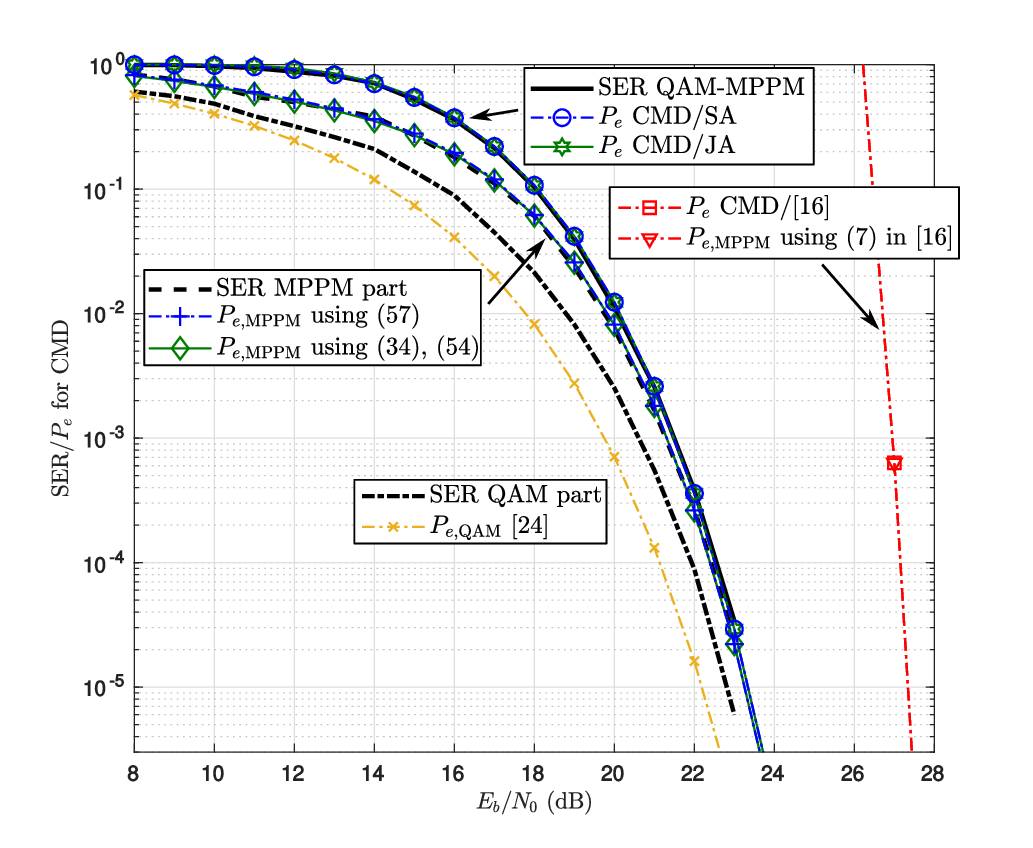}
 \caption{SER and $P_e$ results for the CMD, when $N=12$, $w=6$, $M_Q=16$, $m=0.5$.} \label{Fig2}
\end{figure}

\begin{figure}[htb!]
\centering
\includegraphics[width=\figurewidth]{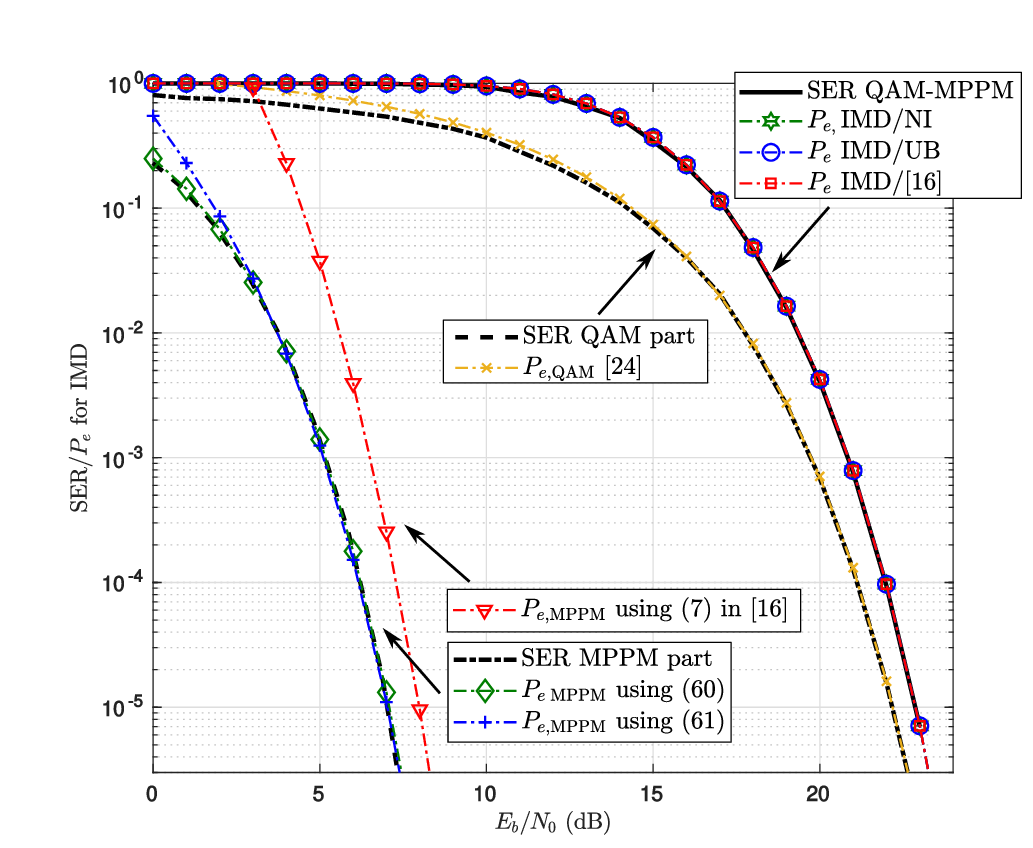}
 \caption{SER and $P_e$ results for the IMD, when $N=12$, $w=6$, $M_Q=16$, $m=0.5$.} \label{Fig3}
\end{figure}
In Fig. \ref{Fig2}, we can see, for the CMD, the symbol error rate (SER) and the average symbol error probability ($P_e$), calculated with different approximations, as a function of $E_b/N_0$, when $N=12$, $w=6$, $M_Q=16$, and $m=0.5$. We can also see the SER of the MPPM part and of the QAM part. It is to be noticed that the average symbol error probabilities given through approximations CMD/JA and CMD/SA are very tight, and their difference is negligible. This means that assumption \eqref{assumption_indep} is really reasonable for the overall symbol error probability. The curve labeled ``$P_e$ using \eqref{PcMPPM-pat} and \eqref{scaled_ver}'' has been calculated averaging over $P_{c,\mathrm{MPPM}}\left(\left\{ \mathbf{s}_{i_j}\right\}_{\mathbf{I}}\right)$ in \eqref{PcMPPM-pat}, resorting to \eqref{scaled_ver} for the numerical calculations task. We can see that it fits the experimental value of MPPM SER very tightly, as well as the average of \eqref{Pc_IMD_SA}. This is no surprise, since both views are formally correct and should lead to the same result, excepting numerical issues. On the other hand, the already published approximation CMD/\cite{6876375} greatly overestimates $P_e$ for QAM-MPPM, due to the fact that the MPPM average symbol error part calculated through (7) in \cite{6876375} is also overestimated.

Notice that the theoretical value of the average symbol error probability for QAM ($P_{e,\mathrm{QAM}}$ from \cite{Pro95}) does not fit the experimental value of the QAM SER. This last value is the counting of all the demodulation QAM symbol errors for the correctly identified signal slots, irrespective whether the whole MPPM symbol is correctly detected or not. The difference between the simulation and the theoretical $P_{e,\mathrm{QAM}}$ is due to the fact that there exists a bias in the QAM symbols that actually enter the QAM demodulator stage: the signal slots corresponding to QAM symbols with higher energy are correctly identified as such with higher probability during the MPPM detection stage than the ones corresponding to QAM symbols with lower energy. The actual difference is small, but this is a proof that the dependence between the MPPM decision stage and the QAM decision stage should be taken into account if we want to make exact calculations. Though not shown, in the CMD/JA case, $P_e$ does not converge to the actual SER value for the lowest signal-to-noise ratio: this is due to the fact that we are using the union bound approximation \eqref{UB_QAM} to account for the individual QAM symbol error probability.

In Fig. \ref{Fig3}, we can see the results for the IMD in a setup with the same parameters as in Fig. \ref{Fig2}. First of all, we may appreciate that there is a gain of around $0.7$ dB when using the IMD with respect to using the CMD. Respecting the three possible approximations for the average symbol error probability $P_e$, we see that they all are very tight. The reason is that the average symbol error probability of QAM, $P_{e,\mathrm{QAM}}$, dominates over the MPPM part, and it is not so important how $P_{e,\mathrm{MPPM}}$ is adjusted. In fact, as it may be seen, the average symbol error probability for MPPM is again overestimated through (7) in \cite{6876375}, but
\begin{equation}
 P_e \approx 1 - \left(1-P_{e,\mathrm{QAM}}\right)^w,
\end{equation}
for $E_b/N_0>10$ dB in all the cases, when $P_e$ starts to fall from $10^0$. The IMD/NI and the IMD/UB scenarios adjust $P_{e,\mathrm{MPPM}}$  even better, but the result in $P_e$ is indistinguishable due to the reasons given. It is to be noted that, in systems where the MPPM SER is not far from the QAM SER, IMD/\cite{6876375} will yield overestimated results for the overall $P_e$ with respect to the actual SER, as it will be made evident in the last figure. On the other hand, considering IMD/UB, the usual divergence of the union bound for low signal-to-noise ratio can be appreciated, while getting a tight result for $E_b/N_0>2$ dB. We can also see how $P_{e,\mathrm{QAM}}$ fits very well to the experimental QAM SER, excepting in the range of low $E_b/N_0$, where the theoretical approximations of \cite{Pro95} slightly diverge. As the detection of the signal slots is made using a metric independent of the specific QAM symbols, now the QAM SER curve does not exhibit the previous bias, as all the possible QAM symbols are equally represented in the QAM detection stage.

\begin{figure}[htb!]
\centering
\includegraphics[width=\figurewidth]{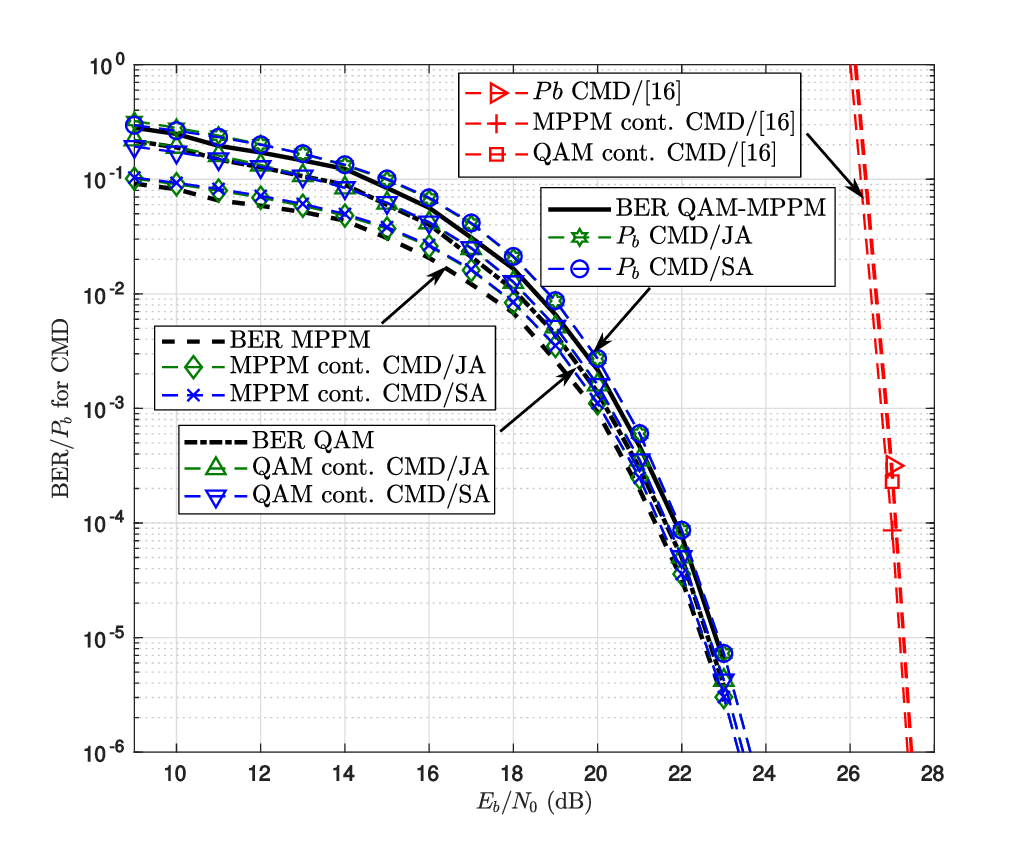}
 \caption{BER and $P_b$ results for the CMD, when $N=12$, $w=6$, $M_Q=16$, $m=0.5$.} \label{Fig4}
\end{figure}
In Fig. \ref{Fig4} we see the average bit error probability and the bit error rate (BER), for the same setup of Fig. \ref{Fig1}. As in the case of the $P_e$, the cases CMD/JA and CMD/SA approximate the final BER with great accuracy, whereas the case CMD/\cite{6876375} results in a very loose upper bound. We also show the different contributions to the $P_b$: the errors associated to the bits in the MPPM part (labeled MPPM cont., and representing the second term on the RHS of \eqref{Pbtot_CMD}), and the errors associated to the bits in the QAM symbols (labeled QAM cont., and representing the first and third term on the RHS of \eqref{Pbtot_CMD}). Again, CMD/JA and CMD/SA methods yield very tight results, whereas the terms in the CMD/\cite{6876375} case are largely overestimated. There is a small mismatch between the MPPM bit error probability computation for CMD/JA and for CMD/SA, because factor \eqref{eq_ne_MPPM} leads to an upper bound approximation \cite{a029b3b4b02e4b40ac8ca72cc4f7fc8c}.

\begin{figure}[htb!]
\centering
\includegraphics[width=\figurewidth]{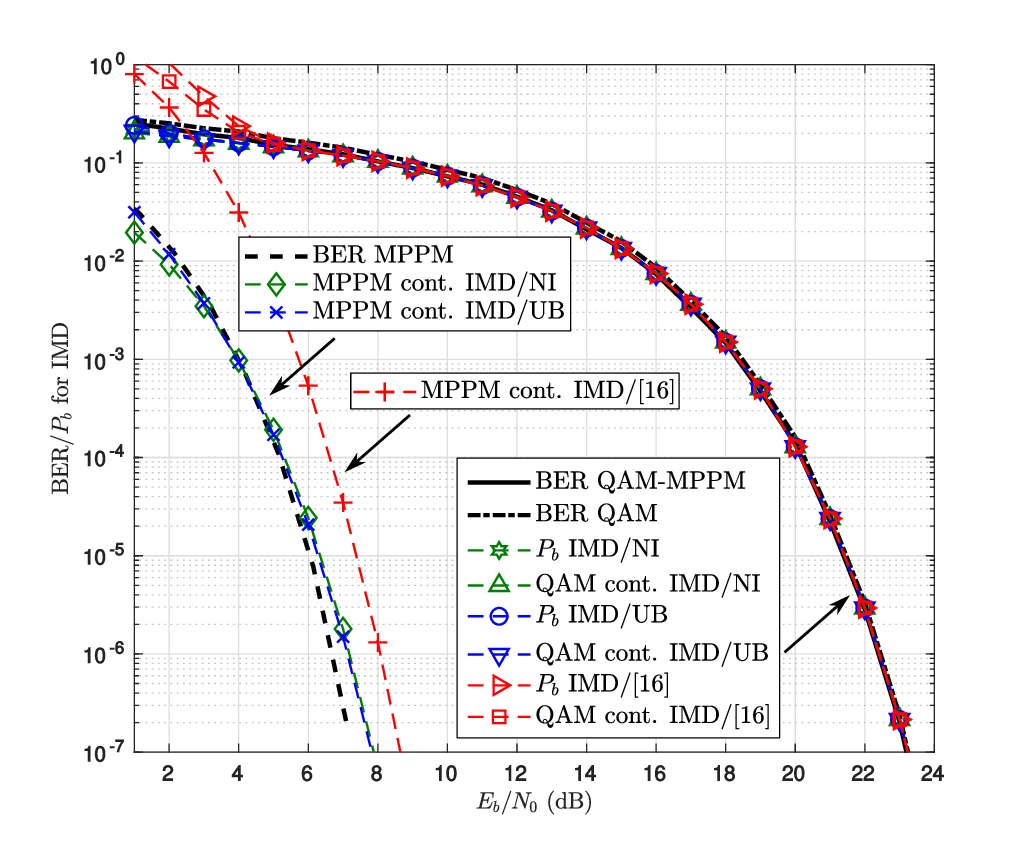}
 \caption{BER and $P_b$ results for the IMD, when $N=12$, $w=6$, $M_Q=16$, $m=0.5$.} \label{Fig5}
\end{figure}
In Fig. \ref{Fig5} we represent the average bit error probability and the BER, for the same setup as in Fig. \ref{Fig3}. The MPPM contribution takes into account the second term on the RHS of \eqref{Pbtot_IMD}, and the QAM contribution the first and third terms thereon. For high $E_b/N_0$, the different approximations yield similar results, though IMD/\cite{6876375} starts diverging. This is due to the already known fact that the MPPM error rate is overestimated: its effect becomes rapidly negligible and $P_{e,\mathrm{QAM}}$, which is estimated in the same way for all the approximations, dominates the $P_b$. In fact, the QAM BER contribution collapses very fast to the QAM-MPPM BER, and so do the approximations. There are only differences in the MPPM contribution, which is slightly different for IMD/NI and IMD/UB: this is due to the inaccuracies of \eqref{eq_ne_MPPM} and of the union bound. Notice also that there is a gain of some tenths of dB in $E_b/N_0$ with respect to the CMD case in Fig. \ref{Fig4}.

\begin{figure}[htb!]
\centering
\includegraphics[width=\figurewidth]{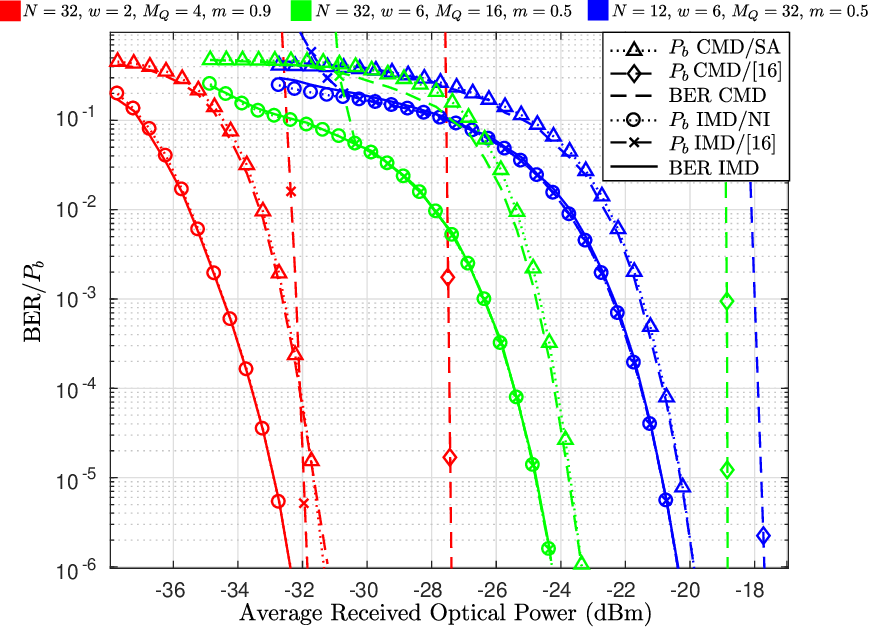}
\caption{BER and $P_b$ results for the IMD and CMD for several cases of interest. The parameters have been chosen to represent a good performing case (red plots) with $N=32$, $w=2$, $M_Q=4$, $m=0.9$, an average case (green plots) with  $N=32$, $w=6$, $M_Q=16$, $m=0.5$, and a poor performing case (blue plots) with $N=12$, $w=6$, $M_Q=16$, $m=0.5$. } \label{Fig6}
\end{figure}
In Fig,  \ref{Fig6}, we represent some results spanning a variety of cases. This time, they are plotted as a function of the received optical power $P_{opt}$, using \eqref{N0} and typical parameter values \cite{7858135}: $T=290$ K, $R_L=50$ $\Omega$, $NF=10\log_{10}\left(F\right)=10$ dB, $(RIN)=-155$ dB/Hz, and $\mathcal{R}=0.5$ A/W. The slot duration $T_s$ has been chosen such that the binary rate is $R_b=50$ Mbps. Apart from the cases CMD/\cite{6876375} and IMD/\cite{6876375}, we have only depicted the approximations corresponding to CMD/SA and IMD/NI, because they give results practically identical to the ones obtained with CMD/JA and IMD/UB, respectively, while being faster in their computations. We can see that the trends identified in the previous figures are kept here: IMD offers a gain with respect to CMD, and the proposed approximations  for CMD are far tighter than the ones presented in \cite{6876375}. Notice that for the case $N=32$, $w=2$, $M_Q=4$ and $m=0.9$, the bound IMD/\cite{6876375} is far less tight than what has been seen in the previous figures because the QAM error part is no longer dominant, and the mismatch in the calculation of the MPPM symbol error probability is made evident.

\section{Conclusions}
\label{conclusions}

In this work we have presented a new method to demodulate an already proposed index modulated waveform intended for the optical channel, called QAM-MPPM. We have derived analytical expressions to calculate the average symbol and bit error probabilities in the AWGN channel, both for the new detector and for the previously published one \cite{6876375}. We have also proposed approximations and practical methods to calculate the analytical values for the average symbol and bit error probabilities, and we have shown through simulation that our proposals are a very good fit for both detectors. We have also verified that there is a gain of some tenths of dB in $E_b/N_0$ when applying the new demodulation method, at practically no additional cost. This is a clear advantage, since the transmitter is the same, and the receiver only has to include a filter matched to the MPPM waveform. The complexity and resources required to detect the MPPM and the QAM bits are similar for both detectors, with a slight advantage for our proposal. We are confident that the principles and methods developed here will help to provide tools to better set and analyze QAM-MPPM in a variety of scenarios, and, as a consequence, to contribute to its practical implementation.


%

%
%
%
%
%

\ifCLASSOPTIONcaptionsoff
  \newpage
\fi



\bibliographystyle{IEEEtran}
\end{document}